\documentclass{IEEEcsmag}

\usepackage[colorlinks,urlcolor=blue,linkcolor=blue,citecolor=blue]{hyperref}

\usepackage{upmath}

\jvol{XX}
\jnum{XX}
\paper{8}
\jmonth{May/June}
\jname{Preprint}
\pubyear{2020}

\usepackage{amssymb}

\usepackage[T1]{fontenc}

\usepackage{setspace}

\begin{document}

\sptitle{Preprint}

\title{A Blockchain Architecture for Industrial Applications}

\author{{L}odovica Marchesi, Michele Marchesi, Roberto Tonelli}
\affil{DMI, University of Cagliari}


\begin{abstract}
The Blockchain and the programs running on it, called Smart Contracts, are more and more applied in all fields requiring trust and strong certifications.
In this work we compare public and permissioned blockchains for industrial applications.
We propose a complete, original solution based on Ethereum to implement a decentralized application.
This solution is characterized by a set of validator nodes running the blockchain using Proof-of-Authority consensus, and including an Explorer enabling users to check blockchain state, and the source code of the Smart Contracts running on it.
From time to time, the hash digest of the last mined block is written into a public blockchain to guarantee immutability.
The right to send transactions is granted by validator nodes to users by endowing them with the local Ethers mined.
Overall, the proposed approach has the same transparency and immutability of a public blockchain, without its drawbacks.
\end{abstract}

\maketitle

\chapterinitial{Introduction}
\label{S:1}
The blockchain technology is a smart mix of known technologies, first introduced in 2008 with the famous paper by Satoshi Nakamoto, as the enabler of the digital currency known as Bitcoin~\cite{nakamoto2008}.
The main requirement of Bitcoin, as of every currency, was trust. 
In fact, to put your saving and the efforts of your work into a currency, you need to trust that the money will be still expendable for a reasonable number of years, secure against counterfeiting, theft, confiscation, double-spending and inflation, and easily transferable.
Another requirement, that diversified Bitcoin from any other currency, was the absence of a central authority managing it.

In Bitcoin, these features were obtained by using a distributed peer-to-peer network, based on open source software running in every node, each holding a copy of the transaction database, known as \textit{blockchain}.
The high number of nodes, and the economic incentives given to miners (who validate transactions and pack them into blocks), guarantees the survival of the network and its robustness; the equality and openness of all nodes makes the network decentralized; the transparency and immutability of the blockchain enables trust; the consensus mechanism used to add new transactions to the blockchain (proof of work) guarantee against massive Sybil attacks; the limited number of Bitcoins mined guarantees against inflation.
The success of Bitcoin, whose market value changed from zero to several thousands of USD in less than ten years, witnesses the success of the Bitcoin vision.

In this paper, however, we are not interested in cryptocurrencies, but in other applications of the blockchain.
In fact, a few years after the introduction of Bitcoin in 2009, managers and developers realized that a blockchain can be also used to run a decentralized computer. 
The first successful network able to run Turing-complete programs, called "Smart Contracts" (SCs) following an idea of Nick Szabo~\cite{szabo1997}, was the Ethereum blockchain, started in 2015~\cite{wood2014}.

\begin{table*}[t] 
\centering
\caption{The features needed by a dApp system.}
\label{T1}
\begin{tabular}{|p{0.5 cm}|p{2.2 cm}|p{11.2 cm}|}
\hline
\multicolumn{1}{|l|}  \# & \textbf{Feature} & \textbf{Description} \\ \hline
1 & No central authority &
If a central authority can run the application, there is no point to use a blockchain!  \\ \hline
2 & Persistency & The system repositories and apps, blockchain included, must be kept runnng for a suitable amount of time -- typically in the range of years or decades -- with negligible risk of being interrupted or terminated before the time. \\ \hline
3 & Immutability & A blockchain is an append-only system -- once written, the information cannot be changed or deleted. The data and programs running on the blockchain must be verifiably immutable and counterfeit proof. \\ \hline
4 & Transparency & The data and the activities performed on the blockchain must be entirely traceable. Anyone, possibly with suitable access rights, should be able to easily explore the blockchain, to verify this. \\ \hline
5 & Parties identification & All writing activities performed on the blockchain must come from certified identities. \\ \hline
6 & Privacy & The permission to access the blockchain, in particular to change its state, must be granted only to known users, possibly at various access levels. \\ \hline
7 & Low, predictable cost & The blockchain system should be open source, easily deployed, and requiring limited hardware and network bandwith resources, compatibly with the size of the dApp, and the number of transactions per second. The cost should not be volatile. \\ \hline
8 & Efficiency & The system should be able to bear the required throughput, with proper response times, even in the case of many users and many transactions per unit of time. \\ \hline
9 & Scalability & The system should be able to scale, if needed. \\ \hline
\end{tabular}
\end{table*}

\section{Uses of dApps and kinds of blockchains}

The software programs using a blockchain are called "decentralized applications", or "dApps", and are one of main new trends of software development. 
They include the SCs actually running on a blockchain, but also the software managing data outside the blockchain and the user interface to interact with it.

Initially, the primary use of SCs -- typically running on Ethereum blockchain -- was to manage second-level digital currencies, called "tokens", mainly used to finance the Initial Coin Offers -- crowfunding operations gathering money in the form of cryptocurrency to finance startups~\cite{fenu2018}.

Besides tokens, dApps are being used for many applications, in the fields of data notarization, finance and insurance contracts, supply chain management, document and workflow management, gambling, gaming, voting, and many others~\cite{xu2019}.
dApps and SCs can be used for automated enforcement of contractual obligations, without having to trust a central authority, and without space and time constraints.

\begin{table*}[t] 
\centering
\caption{Comparison of public and permissioned blockchains.}
\label{T2}
\begin{tabular}{|p{2.2 cm}|p{6 cm}| p{6 cm}|}
\hline
\multicolumn{1}{|l|}{\textbf{Feature}} & \textbf{Public blockchain} & \textbf{Permissioned blockchain} \\ \hline
No central authority & Totally achieved & Achieved, but with much fewer nodes. \\ \hline
Persistency & Very high & High, depending on the willingness and convenience of the validators. \\ \hline
Immutability & 
Very high & 
High, can be very high. \\ \hline
Transparency & 
Very high & 
Depending on the system; can be very high. \\ \hline
Parties identification & 
Very strong, based on private key ownership, if the owner wishes to be identified. & 
Strong if based on username and password, very strong if based on private key ownership. \\ \hline
Privacy & 
Non-existent. & 
Can be enforced. \\ \hline
Low, predictable cost & 
Only software development and execution costs. The latter costs can be very volatile and unpredictable. & 
Infrastructural costs are typically low. Execution costs are low and predictable. \\ \hline
Efficiency & 
The number of transactions per second is quite low. & 
The number of transactions per second can be high. \\ \hline
Scalability & 
Poor scalability if the number of dApps and users increases. & 
High, by deploying further blockchains on the same node, and/or by splitting the nodes. \\ \hline
\end{tabular}
\end{table*}

The features a dApp system must exhibit are reported in Table~\ref{T1}.
The main one is the fact that a central authority running the system is missed. 
The reasons for this may be various. For instance, no single organization might be willing to run the system for cost, or legal liability reasons; or the organizations involved might not wish to let just one of them run the system; or having many, independent nodes could be a guarantee of persistence and immutability of the system.

Features 2-6 are basically aimed to gain user trust without having to trust all blockchain nodes. 
Features 7-9 are desirable for all software systems, but are especially difficult to obtain in public blockchains.

Developing a dApp system, the first issue to address is whether to use a public or a permissioned blockchain.
Public blockchains are open to everyone, and have no central authority.
The most used blockchains for implementing dApps is Ethereum, but others are available, such as EOS, NEO, Qtum and others.

Permissioned blockchains, often called also "distributed ledgers". are managed by a consortium of organizations, which run the validator nodes and perform the consensus mechanism. 
Further validator nodes can be added only by obtaining the permission of existing ones. 
Validators may allow including nodes holding a copy of the blockchain, but unable to validate transactions.
The right to send transactions, and thus to change the state of the blockchain, can be given to users who are not validators.
Also, often everyone can access the blockchain to verify its correctness. 
In other cases, also the read-only access is granted only to users with the permission to do so.

To run a permissioned blockchain, there are many software systems available.
Most public blockchain software is open source, and suitable also to be used to manage a permissioned one.
A prominent permissioned blockchain, aimed mainly to banking applications, is Ripple, developed by a private company and presently run by about 150 invited validators.
There are also projects aimed to build permissioned blockchain software.
The most popular among these is Hyperledger, an open source collaborative effort hosted by the Linux Foundation, aimed to build cross-industry blockchain technologies.

Table~\ref{T2} shows a comparison of public and permissioned blockchains with regard to the properties shown in Table~\ref{T1}.

Public blockchains look the most stable, and easiest to start with, but lack performances and scalability.
Their cost is not predictable, due to the high volatility in cryptocurrencies values and in transaction validation fees.
Moreover, they do not support privacy of data, and are thus non-compliant with respect to the strict guidelines of modern privacy laws, such as European GDPR.

For these reasons, public blockchains are mainly used for applications managing digital money, such as the above cited tokens, and for the notarization of information.
In our proposal, we will focus on non-monetary, industrial applications, and thus on a permissioned blockchain.
However, the proposed solution includes the use of a public blockchain.

\begin{figure*}[ht]
\centering 
\includegraphics[width=14cm]{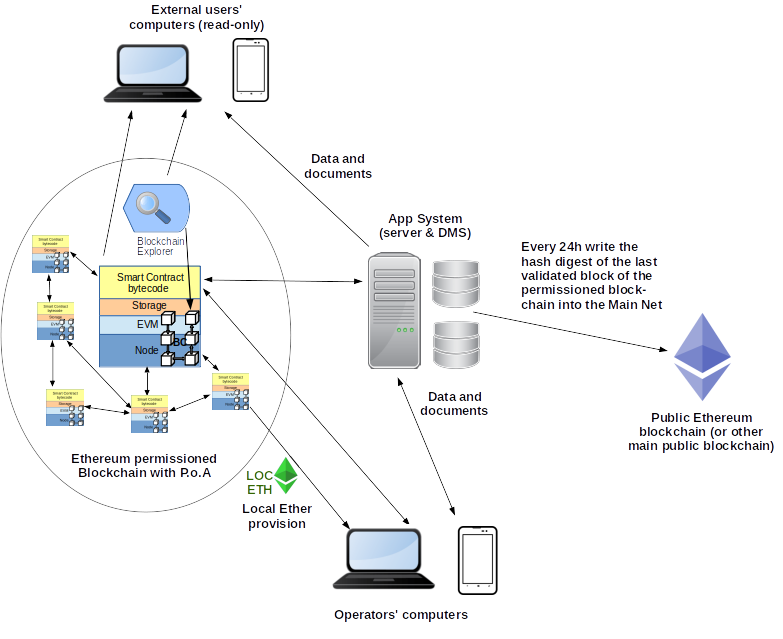}
\caption{The proposed architecture of a dApp application.}
\label{fig:fig1}
\end{figure*}

\section{The proposed dApp architecture}
\label{sec:dapp}

We define as dApp a software system that uses distributed ledger technology (DLT), typically a blockchain, as a central hub to store and exchange information, through Smart Contracts (SCs).

A dApp is composed of SCs running on a blockchain, and of applications able to create and send transactions to SCs. 
This application typically provides a human interaction interface, running on a PC or on a mobile device.
Additional information could be stored on a server, and a business logic could be executed on this.

In our proposal, we will use Ethereum as a reference blokchain, being presently the most used system to develop SCs. 
As of December 2019, 2717 dApps were running on Ethereum public blockchain, out of a total of 3236 surveyed dApps~\cite{sod2019}. 
Another feature of Ethereum is that, to be able to send transactions, users must spend GAS, that is a fraction of Ethers (Ethereum currency) proportional to the resources used.

The proposed dApp architecture is shown in Fig.~\ref{fig:fig1}.
The main component are the P2P validator nodes hosting the blockchain, shown on the left.
These should be managed by independent organizations, to avoid that a single organization might try to falsify the blockchain, or simply decide to stop supporting the dApp. 
We suggest a minimum number of seven independent nodes.

The software running in the nodes is Ethereum, with Proof-of-Authority (PoA) consensus for block validation~\cite{poa2020}.
PoA is quicker and more suitable to a permissioned blockchain than PoW. 
The validators also create Ethers expendable in the permissioned blockchain, which we call "Local Ethers" (LOCETH).

A node, beside the blockchain itself, holds its enabling software, which includes the Ethereum Virtual Machine able to run the SCs.
On top of it there is the SC bytecode, endowed with its permanent data (Storage), which are read from the blockchain for their execution.
All the nodes execute the SCs, and the execution result must be the same for all nodes. 
Hence the impossibility for the SCs to access the external world.
They can access only their data and other SCs stored in the blockchain, which are the same in all nodes.

The other dApp component is a software system running on mobile devices and/or on servers, possibly on the Cloud, exchanging information with users and external systems and devices, which we call "App System".
Its User Interface (UI) typically runs on a Web browser. 
The server component stores data and documents that cannot be stored in the blockchain, and performs business computations. 

The proposed architecture has three kinds of actors:

\begin{itemize}
    \item \textbf{Validators}, that is the nodes running the system and shown in Fig.~\ref{fig:fig1}. They validate transactions and group them in blocks. They also mint LOCETH.
    \item \textbf{Operators}, who are enabled to send transactions which change the blockchain state.
    \item \textbf{External users}, who can access the system nodes in read-only mode.
\end{itemize}

Other key ideas of the proposed architecture are the following:

\begin{itemize} 
    \item One or more blockchain nodes provide an explorer, a software that allows all users to access the blockchain transactions and related addresses, and to inspect the source code of SCs. In this way, the transparency of the permissioned blockchain equals that of a public one.
    \item Every 24 hours, or so, the hash digest of the last block validated in the permissioned blockchain is written into a public blockchain, thus making the former as immutable as the latter.
    \item All users can send "view" queries, which return information from the SCs, without changing the blockchain, and which cost no GAS.
    \item The access control is obtained by validators endowing  operators with LOCETH, that is the GAS enabling sending transactions. Further access control can be performed inside SCs, based on the specific address sending a transaction.
    \item The operators generate their own private keys and corresponding addresses, and declare them to the system. In this way, they have strong digital identification. One can trust that transactions coming from an address actually are sent by the associated identity.
    \item If the dApp must hold large amounts of information, such as documents and images, these documents are stored off-chain, by App System. The hash digest of the document and a link to retrieve it is stored in the blockchain, guaranteeing the date of the document, and its integrity.
\end{itemize}

Together with our spinoff FlossLab ltd we already implemented a prototype of the system, aimed to track the provenance and events of a food supply chain.
In this system, external users are the buyers of a product.
Through a QR code in the label, and an app, they can access the dApp and verify all the transformations and relevant events related to the product, and registered by identified actors of the supply chain. 
The actors are the firms which produced or transformed the product, and the independent certifiers of its quality.
An advanced user can also independently access the permissioned blockchain, and verify the data.

\section{Conclusion and Future Work}
\label{S:6}

The key reason to use a blockchain is trust.
The described approach guarantees the same level of trust and transparency of the public blockchain it is anchored to, allowing much better performances and scalability, and at low, predictable cost.

Presently, we are working with a set of Sardinian institutions to start a permissioned blockchain as described above, with the aim to certify the provenance and quality of local products.





\bibliographystyle{plain}
\bibliography{biblio.bib}







\end{document}